\documentclass{aip-cp}

\usepackage[numbers]{natbib}
\usepackage{rotating}
\usepackage{graphicx}

\newcommand{\physdim}[1]{\hspace{1ex} (\mathrm{#1})}

\begin{document}

\title{What We Know About the $\Lambda(1405)$}

\author[aff1]{Tetsuo Hyodo\corref{cor1}}

\affil[aff1]{Yukawa Institute for Theoretical Physics, Kyoto University, Kyoto 606-8502, Japan}
\corresp[cor1]{Corresponding author: hyodo@yukawa.kyoto-u.ac.jp}

\maketitle

\begin{abstract}
The current status of the $\Lambda(1405)$ resonance in the $\bar{K}N$ scattering is summarized. It is shown that the precise experimental data and the theoretical developments in chiral SU(3) dynamics enable us to quantitatively understand the physics around the $\bar{K}N$ threshold. We present the recent theoretical predictions of the $\pi\Sigma$ spectrum and the investigations of the internal structure of the $\Lambda(1405)$.

\end{abstract}

\section{INTRODUCTION}

The $\Lambda(1405)$ is a resonance in the $\pi\Sigma$ scattering with spin-parity $J^{P}=1/2^{-}$, strangeness $S=-1$, and isospin $I=0$. The resonance is found slightly below the $\bar{K}N$ threshold. The nature of the $\Lambda(1405)$ has been under a long-standing debate ever since the $\Lambda(1405)$ was predicted and discovered more than fifty years ago~\cite{Dalitz:1959dn,Dalitz:1960du,Alston:1961zz} (see Ref.~\cite{Hyodo:2011ur} for a review). The failure of the simple three-quark picture to reproduce $\Lambda(1405)$~\cite{Isgur:1978xj} urges people to consider more exotic configurations such as the meson-baryon molecular structure~\cite{Dalitz:1967fp}\footnote{Perhaps it might be instructive to further comment on the difficulty of the ``good-old'' three-quark models of $\Lambda(1405)$. (1) In the heavy quarkonium sector, a simple quark potential model well reproduces the experimental spectra of quarkonia below the $D\bar{D}$/$B\bar{B}$ threshold. However, above the open charm/bottom threshold, the experimental spectra significantly disagree with the prediction and there appears plenty of unpredicted states called ``$XYZ$''~\cite{Brambilla:2010cs} which are considered to be the candidate of the exotic hadrons: multiquark states, hadronic molecules, and gluon hybrids. The $\Lambda(1405)$ appears near the $\bar{K}N$ threshold and decays into $\pi\Sigma$. This is analogous to $X(3872)$~\cite{Choi:2003ue} and $Z_{c}(3900)$~\cite{Ablikim:2013mio,Liu:2013dau} which are found around the $D\bar{D}^{*}$ threshold and decays into the pion(s) and the charmonium. (2) Importance of the hadron-hadron components is discussed also in the light quark sector. Recent comprehensive study of the $N(1440)$ resonance in the $\pi N$ scattering shows that the ``bare'' state of the $N(1440)$, corresponding to the three-quark component, is found around 1800 MeV~\cite{Suzuki:2009nj}. It is therefore dangerous to naively compare the observed mass of excited baryons with the prediction of the static quark model, because of the dynamical meson-baryon effect. This reinforces the importance of the meson-baryon component in the study of the excited baryons. (3) The first principle lattice QCD supports the $\bar{K}N$ molecular structure of $\Lambda(1405)$~\cite{Hall:2014uca}. In this way, there many issues which are inconvenient to the simple three-quark model. To insist the three-quark picture for $\Lambda(1405)$, one has to overcome all these issues and construct a theoretical framework with phenomenologically successful description of the scattering data at the level of $\chi^{2}/\rm{dof}\sim 1$ as achieved in chiral SU(3) dynamics.}. 

From modern viewpoint, the $\Lambda(1405)$ is well described as a resonance state in the meson-baryon scattering in the framework of chiral SU(3) dynamics~\cite{Kaiser:1995eg,Oset:1998it,Oller:2000fj,Lutz:2001yb,Hyodo:2011ur}. In fact, the $\bar{K}N$ system is closely related to the spontaneous and explicit breaking of chiral SU(3)$_{R}\times$SU(3)$_{L}$ symmetry in QCD. The anti-kaon is regarded as the Nambu-Goldstone (NG) boson associated with the spontaneous breaking of chiral symmetry, while the explicit symmetry breaking by the strange quark mass makes the anti-kaon as massive as $\sim 500$ MeV. The interesting dynamics of the $\bar{K}N$ system and the $\Lambda(1405)$ reflects the unique feature of the anti-kaon in hadron physics. 

In recent years, there have been remarkable achievements in the experimental investigations of the $\Lambda(1405)$ and the $\bar{K}N$ system. The SIDDHARTA collaboration has determined the shift $\Delta E$ and width $\Gamma$ of the 1s level of the kaonic hydrogen through the X-ray measurement as~\cite{Bazzi:2011zj,Bazzi:2012eq}
\begin{eqnarray}
\Delta E = 283\pm36(\textit{stat})\pm6(\textit{syst}) \textrm{ eV} , \quad \Gamma = 541\pm89(\textit{stat})\pm22(\textit{syst})\textrm{ eV} . \label{eq:SIDDHARTA} 
\end{eqnarray}
Because the shift and width can be related to the $K^{-}p$ scattering length~\cite{Meissner:2004jr}, this experimental data provides a strong and direct constraint on the $\bar{K}N$ scattering amplitude at the fixed energy. At the same time, the $\Lambda(1405)$ signal in the $\pi\Sigma$ spectra has been observed in various production experiments, such as the photoproductions by the LEPS collaboration~\cite{Niiyama:2008rt} and by the CLAS collaboration~\cite{Moriya:2013eb,Moriya:2013hwg}, and the proton-proton collisions by the HADES collaboration~\cite{Agakishiev:2012xk}. These spectra are qualitatively different from the old one~\cite{Hemingway:1985pz} where only a single $\pi^{-}\Sigma^{+}$ spectrum is extracted without the normalization of the magnitude. In the new data~\cite{Moriya:2013eb}, the absolute value of the differential cross sections are determined in all three charge combinations, $\pi^{\pm}\Sigma^{\mp}$ and $\pi^{0}\Sigma^{0}$. This experimentally verifies the isospin interference effect in the $\pi\Sigma$ spectrum discussed in Ref.~\cite{Nacher:1998mi}. The precise measurement also leads to the experimental determination of the spin-parity of the $\Lambda(1405)$~\cite{Moriya:2014kpv}. 

These experimental developments have stimulated the theoretical studies. In the following, we first discuss the meson-baryon coupled-channel scattering with the $\Lambda(1405)$ in the framework of the chiral SU(3) dynamics. The systematic improvement with the next-to-leading order (NLO) interaction is now possible owing to the constraint by the SIDDHARTA data~(\ref{eq:SIDDHARTA}). We then consider the $\pi\Sigma$ spectrum in which the $\Lambda(1405)$ signal is observed. Finally, we discuss the structure of the $\Lambda(1405)$ from the viewpoint of the $\bar{K}N$ wavefunction and the compositeness.

\section{CHIRAL SU(3) DYNAMICS}

\subsection{$\Lambda(1405)$ and $\bar{K}N$ scattering}

The interaction of the NG bosons with hadrons is dictated by the chiral low-energy theorems, which can be systematically improved by chiral perturbation theory~\cite{Scherer:2012xha}. According to the low-energy theorem~\cite{Weinberg:1966kf,Tomozawa:1966jm}, the $s$-wave scattering of the NG boson with any target hadron is proportional to the energy of the NG boson. The explicit symmetry breaking provides the mass of the NG boson, which enhances the interaction strength. For pions, the effect of the explicit breaking is considered to be small, while for kaons, the enhancement of the interaction by the strange quark mass becomes significant. Thus, in contrast to the $\pi\pi$ and $\pi N$ scatterings, the low-energy interaction of the $\bar{K}N$ system is not perturbatively small. This is consistent with the existence of the $\Lambda(1405)$ resonance below the $\bar{K}N$ threshold, which prevents the description by the perturbative calculation. We thus need to sum up the chiral interaction $V(W)$ nonperturbatively to obtain the scattering amplitude $T(W)$ as
\begin{eqnarray}
    T(W) = V(W)+V(W)G(W)T(W) ,
   \label{eq:amplitude} 
\end{eqnarray}
with the loop function $G(W)$. The $\Lambda(1405)$ and the coupled-channel meson-baryon scattering are well described in this approach~\cite{Kaiser:1995eg,Oset:1998it,Oller:2000fj,Lutz:2001yb,Hyodo:2011ur}. This framework is similar to the chiral effective field theory approach for nuclear force~\cite{Epelbaum:2008ga,Machleidt:2011zz}; the interaction potential is constructed by the chiral perturbation theory and the full amplitude is obtained by solving the scattering equation.

\begin{figure}[tb]
  \centerline{
  \includegraphics[width=0.5\textwidth,bb=-95 0 430 425]{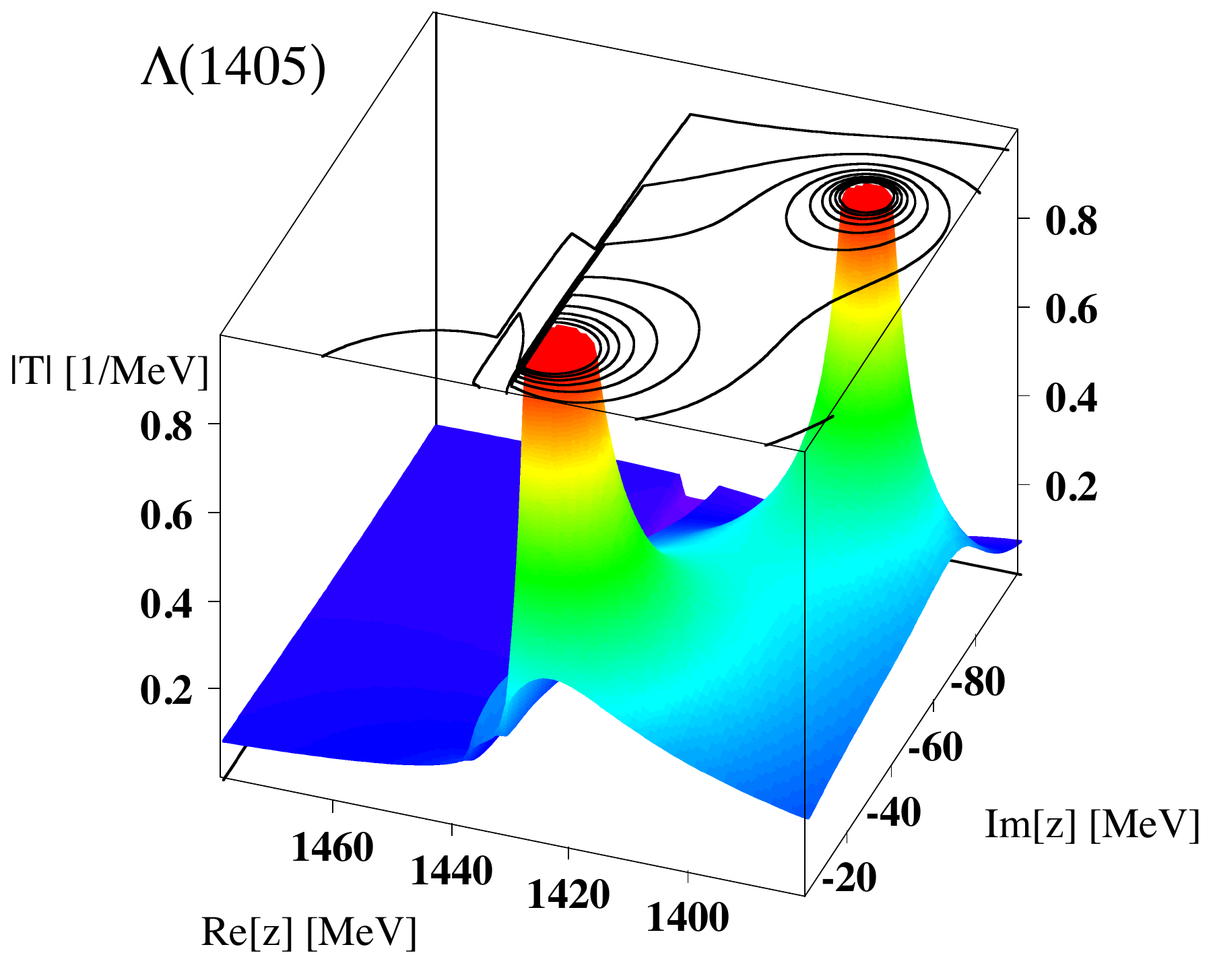}
  }
  \caption{Absolute value of the scattering amplitude $|T(z)|$ of the $\bar{K}N$ elastic channel in the second Riemann sheet of the complex energy $z$ plane. The figure is taken from Ref.~\cite{Hyodo:2011ur}.}
  \label{fig:pole1405}
\end{figure}

The pole structure of the $\Lambda(1405)$ in the complex energy plane shows an interesting property. In general, a resonance is associated with the pole singularity of the scattering amplitude, corresponding to the eigenenergy of the Hamiltonian. It has been found that the $\Lambda(1405)$ is associated with two poles~\cite{Oller:2000fj,Jido:2003cb}, as shown in Fig.~\ref{fig:pole1405}. The origin of two poles is understood by the two attractive interactions in this sector: flavor singlet and octet channels in the SU(3) basis~\cite{Jido:2003cb} and the $\bar{K}N$ and $\pi\Sigma$ channels in the isospin basis~\cite{Hyodo:2007jq}. Two poles can be interpreted as the $\bar{K}N$ quasi-bound state and the $\pi\Sigma$ resonance by these attractive interactions. The existence of the attractions is guaranteed by the low-energy theorem.

\subsection{Analysis with precise kaonic hydrogen data}

It is the virtue of the chiral SU(3) dynamics that the theory can be systematically improved by introducing the higher order contributions. Of course, the higher order terms contain the low-energy constants which need to be determined from the comparison with the experimental data. In other words, the improvement by the higher order terms is meaningful only when the sufficient amount of reliable experimental data is available. There have been several studies with higher order terms before the SIDDHARTA data~\cite{Kaiser:1995eg,Lutz:2001yb,Borasoy:2004kk,Borasoy:2005ie,Oller:2005ig,Oller:2006jw,Borasoy:2006sr}, but the uncertainty of the subthreshold extrapolation is not small, as demonstrated in Ref.~\cite{Borasoy:2006sr}. 

\begin{figure}[tb]
  \centerline{
  \includegraphics[width=8cm,bb=0 0 1024 768]{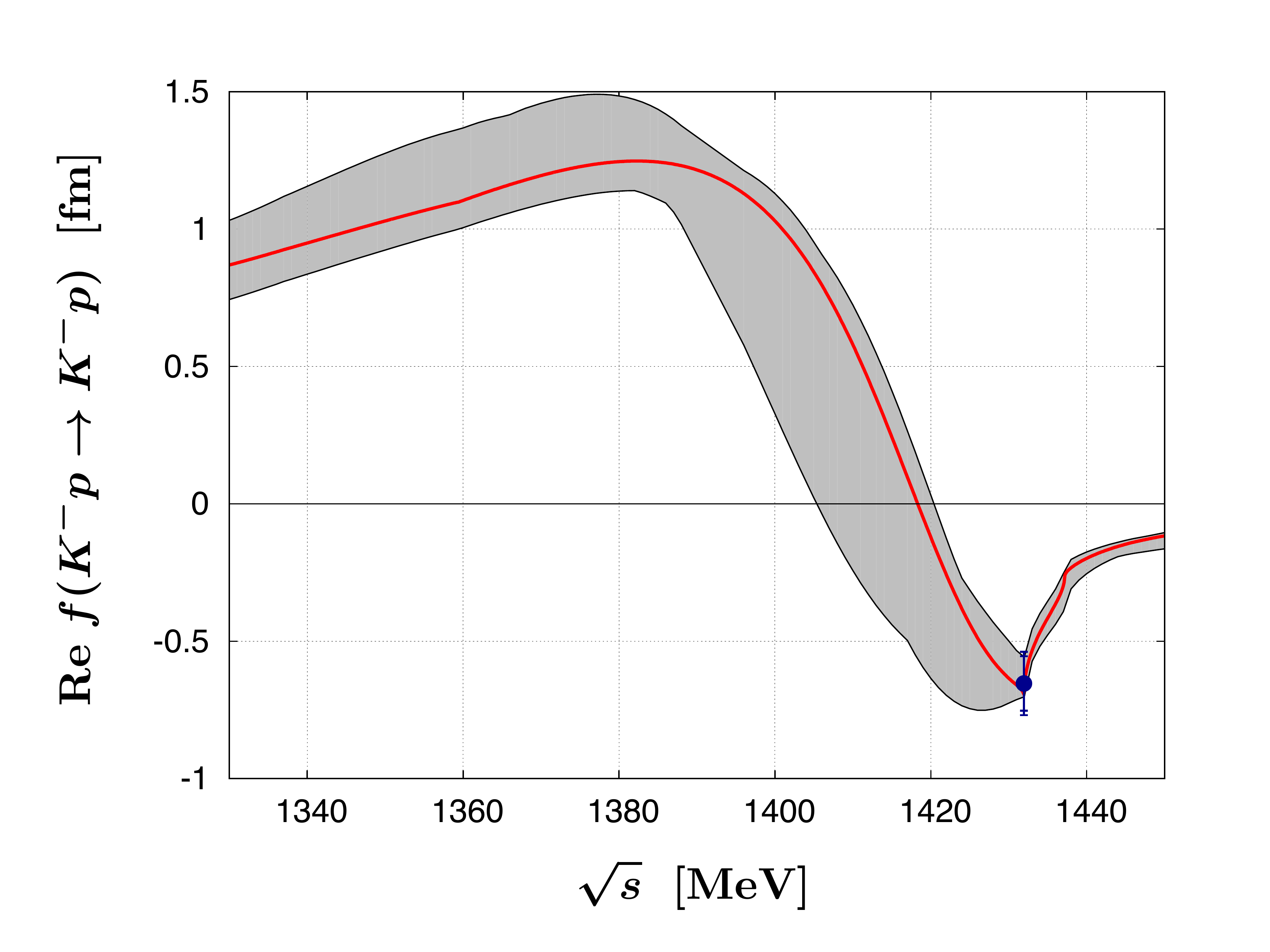}
  \includegraphics[width=8cm,bb=0 0 1024 768]{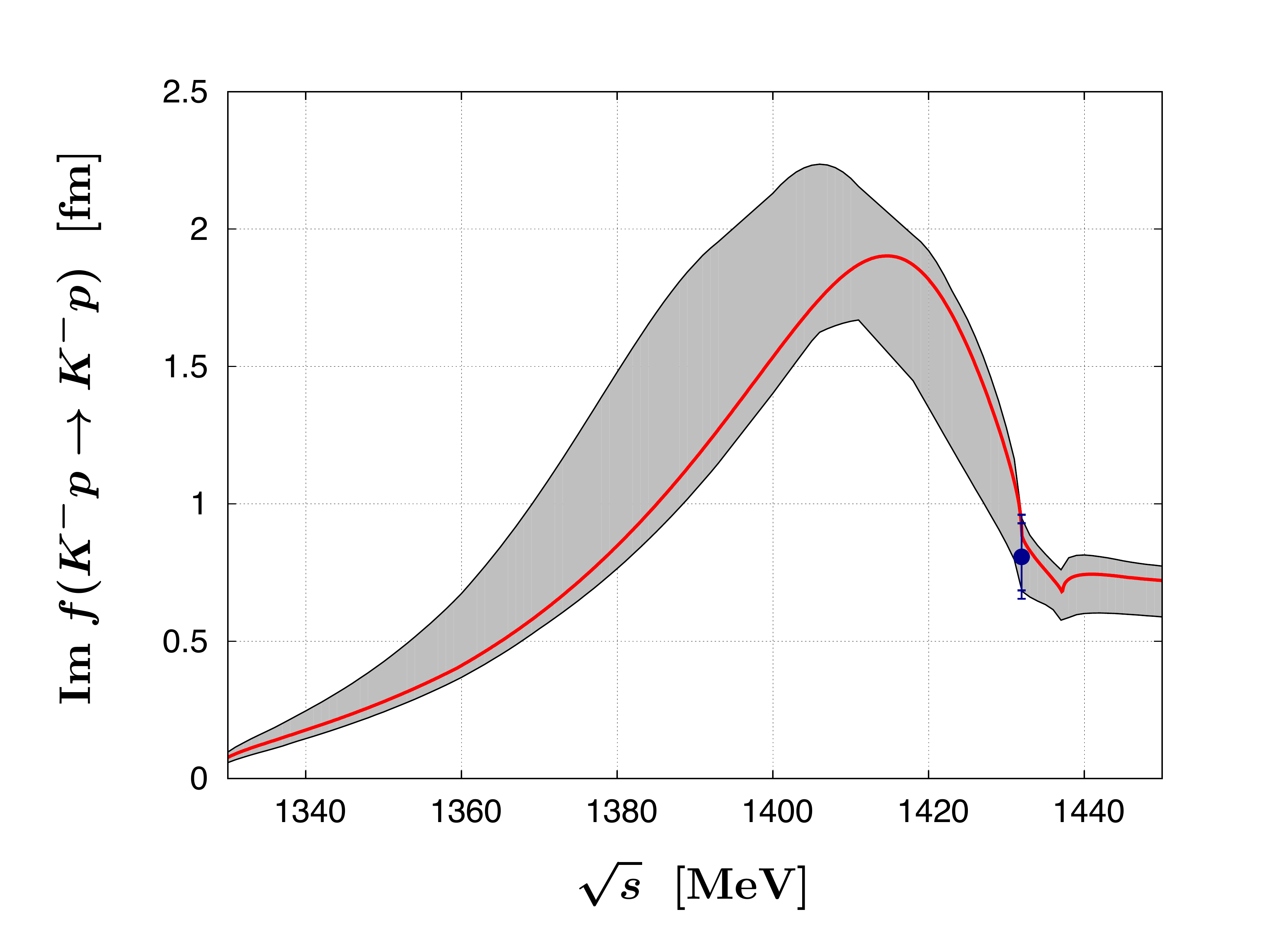}
  }
  \caption{Real part (left) and imaginary part (right) of the $K^-p \rightarrow K^-p$ forward scattering amplitude
extrapolated to the subthreshold region. The shaded bands show the uncertainty from the constraints by the SIDDHARTA data indicated by the dots. The figures are taken from Ref.~\cite{Ikeda:2012au}.}
  \label{fig:ampligude}
\end{figure}

Since the $K^{-}p$ scattering length is precisely determined by SIDDHARTA collaboration~\cite{Bazzi:2011zj,Bazzi:2012eq}, it is now possible to incorporate the NLO terms with this constraint. The systematic NLO analysis is performed in Refs.~\cite{Ikeda:2011pi,Ikeda:2012au} by fitting the $K^{-}p$ total cross sections, threshold branching ratios, and the $K^{-}p$ scattering length by SIDDHARTA. The best fit result is obtained with $\chi^{2}/$dof $=0.96$. We thus find a good description of both the scattering data and the kaonic hydrogen measurement, and solve the inconsistency problem found in the previous measurement of the kaonic hydrogen~\cite{Borasoy:2004kk,Borasoy:2005ie}. In Fig.~\ref{fig:ampligude}, the subthreshold extrapolation of the $\bar{K}N$ amplitude is shown. The uncertainty in the subthreshold extrapolation is significantly reduced, in comparison with the previous studies. The result demonstrates the importance of the precise determination of the $K^{-}p$ scattering length for the subthreshold $\bar{K}N$ amplitude where the $\Lambda(1405)$ exists. Once the amplitude is determined on the real axis, we can extrapolate it to the complex energy plane. Two poles are found in the energy region of the $\Lambda(1405)$ as
\begin{eqnarray}
    z_{1} =1424^{+7}_{-23}- i 26^{+3}_{-14} \rm{MeV},
    \quad 
    z_{2} = 1381^{+18}_{-6}- i 81^{+19}_{-8}\rm{MeV} .
\end{eqnarray}
This confirms the two-pole structure of the $\Lambda(1405)$ in the NLO analysis. The pole $z_{1}$ appears around 1420 MeV, not in the nominal position of 1405 MeV. The position of this pole is closely related with the strength of the $\bar{K}N$ interaction~\cite{Hyodo:2007jq}.

Using the same framework with the $K^{-}p$ scattering, it is also possible to predict the $K^{-}n$ scattering lengths which reflects the $I=1$ component of the $s$-wave meson-baryon scattering. The predicted $K^{-}n$ scattering length shows certain amount of deviation with respect to the uncertainty of the input, in comparison with the $I=0$ counterpart. This shows that the $I=1$ component of the scattering length still suffers from the uncertainty. The $K^{-}n$ scattering length can in principle be determined by the kaonic deuterium measurement.

\subsection{Compilation of different approaches}

After Refs.~\cite{Ikeda:2011pi,Ikeda:2012au}, several other groups study the $\Lambda(1405)$ in similar approaches~\cite{Mai:2012dt,Guo:2012vv,Mai:2014xna}. Although the formulation of the nonperturbative scattering amplitude and the strategy of the fitting procedure are different, these studies  commonly use the chiral NLO interaction and perform the $\chi^{2}$ uncertainty analysis including the SIDDHARTA data. Thus, the difference of the results can be understood as the systematic uncertainties of the chiral SU(3) dynamics. We note that an unphysical solution may be found in the fitting procedure~\cite{Mai:2012dt}, reflecting the large number of parameters. It is shown in Ref.~\cite{Mai:2014xna} that the consistency check with the $\pi\Sigma$ spectra excludes the unphysical solution. 

In all works, two poles are found in the $\Lambda(1405)$ energy region, as summarized in Table~\ref{tab:comparison}. As long as the whole bunch of experimental data is respected, the pole 1 is consistently found around 1420 MeV. The same conclusion is obtained by the comprehensive study of the photoproduction spectra with the leading order chiral interaction in Refs.~\cite{Roca:2013av,Roca:2013cca}. On the other hand, the position of the pole 2 does not well converge, because the location of the pole is far from the $\bar{K}N$ threshold. Future study is needed to pin down the precise position of the second pole of the $\Lambda(1405)$.

\begin{table}[bt]	
\caption{
Comparison of the pole positions of $\Lambda(1405)$ in the
complex energy plane from next-to-leading order chiral SU(3) dynamics
including the SIDDHARTA constraint. \label{tab:comparison}
}
\vspace{0.2cm}
\begin{tabular}{lll}
\hline
approach & pole 1 (MeV) & pole 2 (MeV) \\ \hline
Ref.~\cite{Ikeda:2011pi,Ikeda:2012au} NLO 
  & $1424^{+7}_{-23}- i 26^{+3}_{-14}$
  &  $1381^{+18}_{-6}- i 81^{+19}_{-8}$ \\
Ref.~\cite{Guo:2012vv} Fit I 
  & $1417^{+4}_{-4}- i 24^{+7}_{-4}$ 
  & $1436^{+14}_{-10}- i 126^{+24}_{-28}$ \\ 
Ref.~\cite{Guo:2012vv} Fit II 
  & $1421^{+3}_{-2}- i 19^{+8}_{-5}$ 
  & $1388^{+9}_{-9}- i 114^{+24}_{-25}$ \\
Ref.~\cite{Mai:2014xna} solution \#2
  & $1434^{+2}_{-2} - i \, 10^{+2}_{-1}$ 
  & $1330^{+4~}_{-5~} - i \, 56^{+17}_{-11}$\\
Ref.~\cite{Mai:2014xna} solution \#4
  & $1429^{+8}_{-7} - i \, 12^{+2}_{-3}$ 
  & $1325^{+15}_{-15} - i \, 90^{+12}_{-18}$\\
\hline
\end{tabular}
\end{table}

\section{$\Lambda(1405)$ IN $\pi\Sigma$ SPECTRUM}

The $\pi\Sigma$ spectrum is the only accessible experimental information in the energy region of the $\Lambda(1405)$. We however note that the $\pi\Sigma$ spectrum does not provide a \textit{direct} constraint to the meson-baryon scattering amplitude. Except for the (currently impossible) $\pi\Sigma$ elastic scattering experiment, the production of the $\Lambda(1405)$ necessarily involves the initial and final state interactions, on top of the $S=-1$ meson-baryon scattering amplitude. Therefore, a careful analysis is required for each reaction. Here we introduce two recent studies to reveal the effect of the $\Lambda(1405)$ in the $\pi\Sigma$ spectrum.

\subsection{$K^{-}d\to n\pi\Sigma$ reaction with full three-body dynamics}

The $\Lambda(1405)$ production in the $K^{-}d\to n\pi\Sigma$ reaction is now ongoing by the J-PARC E31 collaboration with the kaon incident momentum of about 1 GeV. It is planned to observe all charge combinations of the $\pi\Sigma$ states, which is important to separate the different isospin components. This reaction has been studied in simplified two-step approaches in Refs.~\cite{Jido:2009jf,Miyagawa:2012xz,Jido:2012cy,YamagataSekihara:2012yv}。

In Ref.~\cite{Ohnishi:2015iaq}, we study the $K^{-}d\to n\pi\Sigma$ reaction where all possible correlations are included by the full three-body calculation~\cite{Ohnishi:2013rix}. The relativistic kinematics is employed to deal with the high incident momentum. Examining two $\bar{K}N$ interaction models, we find that the $\pi\Sigma$ spectra are sensitive to the subthreshold behavior of the two-body input. The comparison of the $\pi\Sigma$ spectrum with the forthcoming data by J-PARC E31 experiment will provide further information of the subthreshold $\bar{K}N$ interaction and the $\Lambda(1405)$.

\subsection{Nonleptonic weak decay of $\Lambda_{c}$}

In general, the $\pi\Sigma$ spectrum contains the contamination and the interference with the $I=1$ component, as observed in the photoproduction experiments~\cite{Moriya:2013eb}. Interestingly, Ref.~\cite{Miyahara:2015cja} finds that the nonleptonic weak decay of $\Lambda_{c}\to \pi^{+}MB$ will be an ideal process in this respect. By examining the dominant weak decay process, it is shown that the weak decay produces the meson-baryon pair in the combination of 
\begin{equation}
|MB\rangle = |K^-p\rangle + |\bar{K}^0n\rangle -\frac{\sqrt{2}}{3}|\eta\Lambda\rangle.  \label{eq:hadronstate} 
\end{equation}
This state is in pure $I=0$. Thus, if the dominance of the decay process considered is valid, the weak $\Lambda_{c}$ decay can be regarded as an $I=0$ filter, up to the isospin breaking corrections.

\begin{figure}[tb]
  \centerline{
  \includegraphics[width=8cm,bb=0 0 846 594]{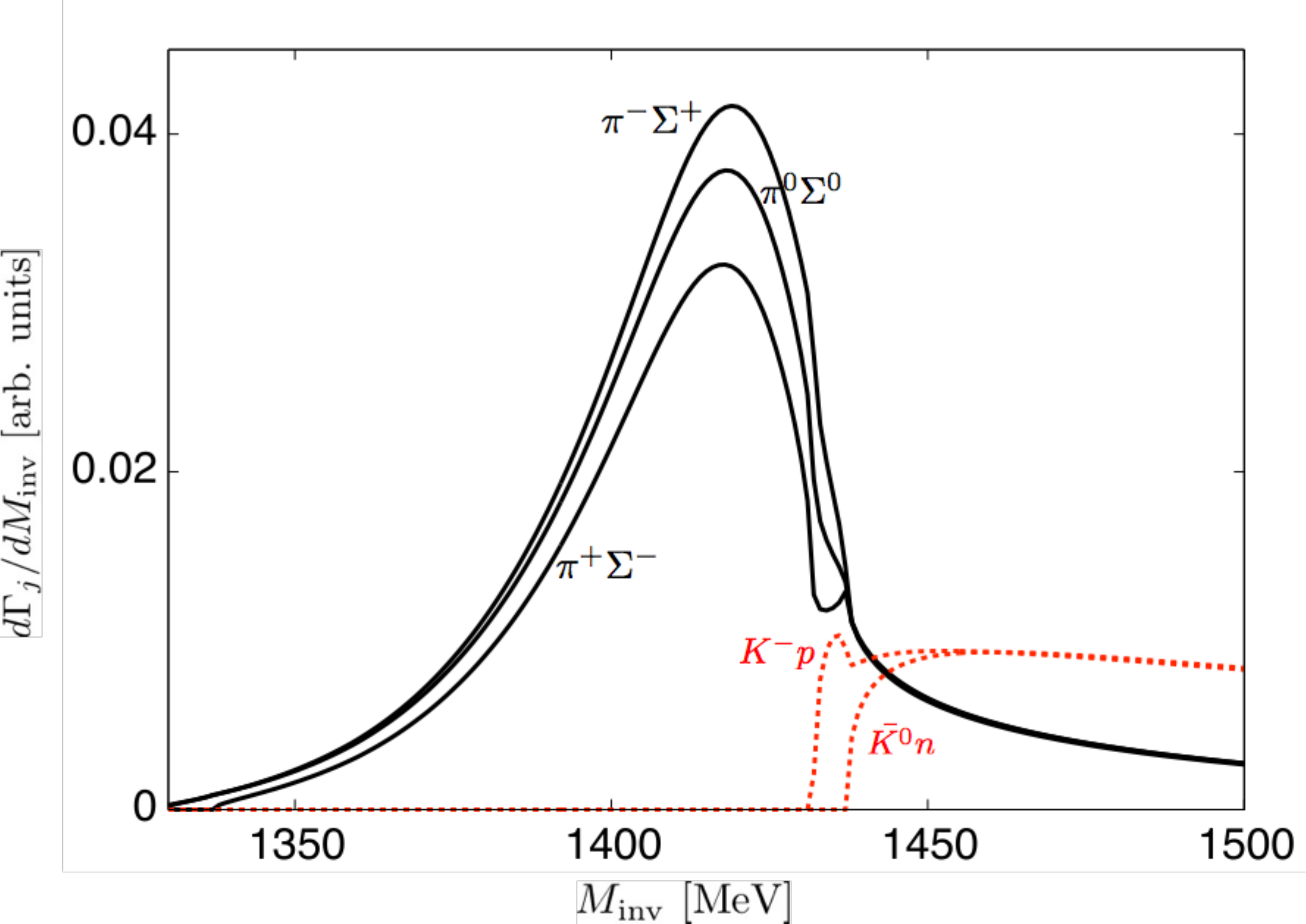}
  }
  \caption{Invariant mass distribution of the decay $\Lambda_c^+\to \pi^+ MB$ near the $\bar{K}N$ threshold with the meson-baryon scattering amplitude in Refs.~\cite{Ikeda:2011pi,Ikeda:2012au}. The solid line represents the spectrum for $\pi\Sigma$ channels and the dashed line for $\bar{K}N$ channels. The figure is taken from Ref.~\cite{Miyahara:2015cja}.}
  \label{fig:spectrum}
\end{figure}

The predicted $\pi\Sigma$ and $\bar{K}N$ spectra are shown in Fig.~\ref{fig:spectrum}, with the meson-baryon scattering amplitude in Refs.~\cite{Ikeda:2011pi,Ikeda:2012au} being employed to represent the final state interaction. Thanks to the isospin filter mechanism, the peak position of the $\Lambda(1405)$ in the $\pi\Sigma$ spectrum does not depend on the charge combination. This suggests that the clean $\Lambda(1405)$ signal can be observed even from the charged $\pi^{\pm}\Sigma^{\mp}$ pairs, which are in general advantageous in the experimental detection. The $\Lambda(1405)$ production in the nonleptonic $\Lambda_{c}$ decay can be studied at Belle and BESIII where the detailed analysis of these decay channels have been recently performed~\cite{Zupanc:2013iki,Ablikim:2015flg}.

\section{STRUCTURE OF $\Lambda(1405)$}

Here we discuss the structure of $\Lambda(1405)$. It should be emphasized  that the success of the chiral SU(3) dynamics within the meson-baryon model space does \textit{not} necessarily indicate the dominance of the meson-baryon component. In general, model parameters (such as cutoff) can represent the effects which are not included in the model space. This mechanism is explicitly demonstrated in Ref.~\cite{Hyodo:2008xr}; the $N(1535)$ resonance is shown to have substantial $qqq$ component, even though the states can be well described by the chiral SU(3) dynamics with the meson-baryon model space. Thus, further efforts are needed to pin down the structure of resonances. In the following, we show two recent studies on the structure fo $\Lambda(1405)$, based on the analysis including SIDDHARTA constraint.\footnote{As discussed by R.L. Jaffe in this conference, the response to the change of the number of colors $N_{c}$ is also useful to study the structure of hadrons. The $N_{c}$ scaling of the $\Lambda(1405)$ is discussed in Refs.~\cite{Hyodo:2007np,Roca:2008kr} where the non-$qqq$ behavior is observed.}

\subsection{Realistic $\bar{K}N$ potential and wavefunction of $\Lambda(1405)$}

Traditionally, the structure of a state is investigated from its wavefunction. Although the chiral SU(3) dynamics does not directly provide the wavefunction of $\Lambda(1405)$, a method to construct an equivalent single-channel potential is established~\cite{Hyodo:2007jq}. The potential is constructed in the local form with complex energy-dependent strength. The local potential can be used to study the property of the two-body quasi-bound state, the $\Lambda(1405)$. 

In Ref.~\cite{Miyahara:2015bya}, the $\bar{K}N$ local potential is constructed, based on the scattering amplitude in the NLO chiral SU(3) dynamics~\cite{Ikeda:2011pi,Ikeda:2012au}. Because this potential reproduces the $\bar{K}N$ scattering amplitude at the level of $\chi^{2}/\rm{dof}\sim 1$, it is reasonable to call it a realistic $\bar{K}N$ potential, in the same spirit with the nuclear force. Solving the Schr\"odinger equation at the pole energy, we obtain the wave function of the $\Lambda(1405)$, $\psi(r)$.

\begin{figure}[tb]
  \centerline{
  \includegraphics[width=8cm,bb=0 0 765 519]{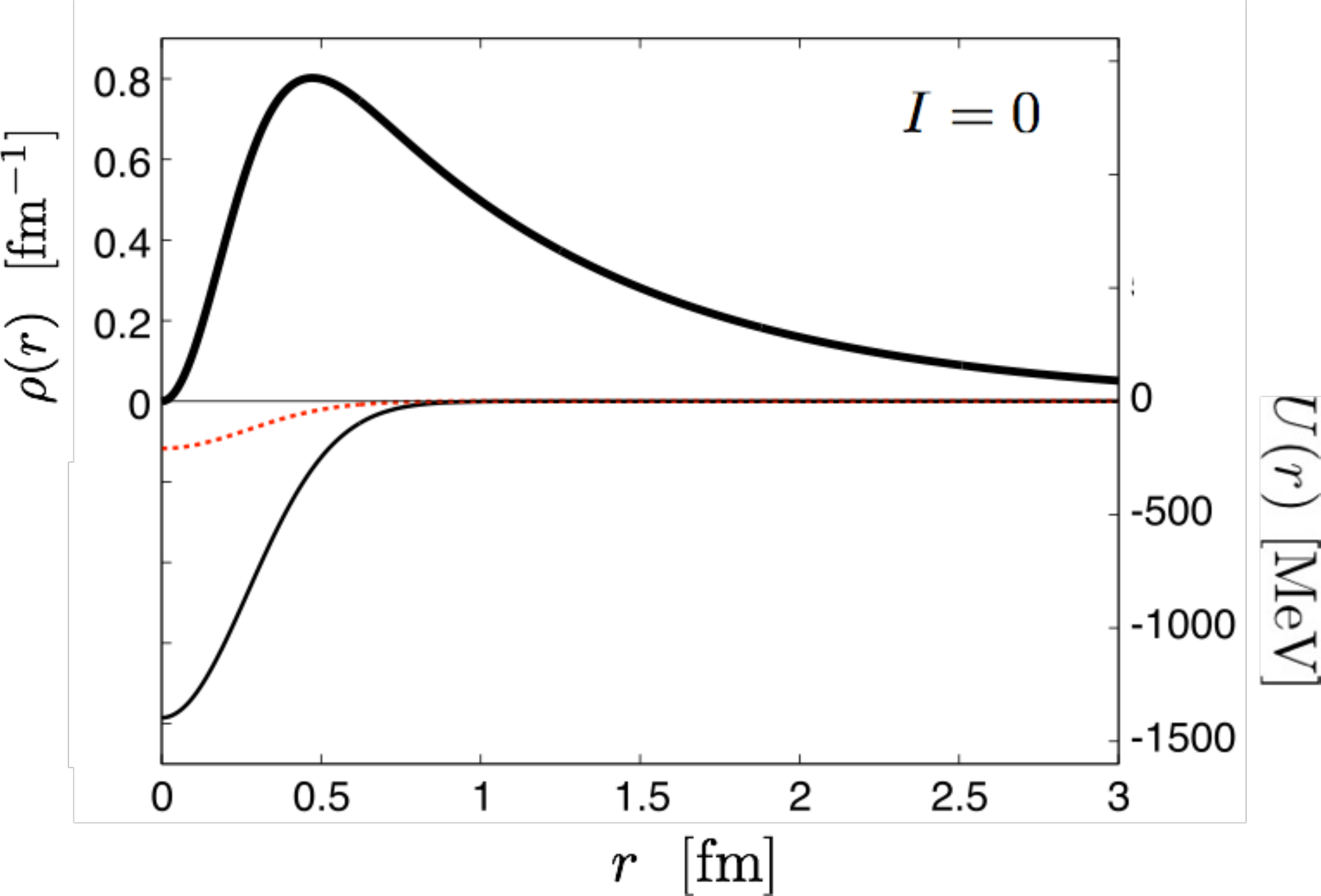}
  }
  \caption{$\bar{K}N$ density distribution $\rho(r)$ (thick solid line), the real part (thin solid line) and the imaginary part (dotted line) of the $\bar{K}N$ potential $(I=0)$ at the $\Lambda(1405)$ pole energy. The figure is taken from Ref.~\cite{Miyahara:2015bya}.}
  \label{fig:potential}
\end{figure}

In Fig.~\ref{fig:potential}, we show the density distribution $\rho(r) = r^{2}|\psi(r)|^{2}$ and the real and imaginary parts of the potential $U(r)$ at the pole energy $1423.97 - 26.28i$ MeV. Because the pole of the $\Lambda(1405)$ is in the first Riemann sheet of the $\bar{K}N$ channel, the wavefunction decreases at large distance. We observe that the $\bar{K}N$ density is distributed in the region $r>1$ fm, where the potential is already vanished. This means that a substantial amount of the $\bar{K}N$ wavefunction exists far outside the potential range. To quantify this fact, we calculate the root mean distance as
\begin{equation}
\sqrt{\langle r^{2}\rangle}
=  \left(
\int d^{3}r \rho(r)
\right)^{1/2}=1.44 \phantom{0}\rm{fm}.
\end{equation}
This is certainly larger than the size of typical hadrons $\sim 0.7$ fm. Thus, the analysis of the wavefunction of the $\Lambda(1405)$ supports its $\bar{K}N$ molecular picture.

\subsection{Compositeness of $\Lambda(1405)$}

It has been shown in Ref.~\cite{Weinberg:1965zz} that the structure of a weakly binding $s$-wave state can be determined even without knowing the detailed structure of the wavefunction and the short range interaction. The weak-binding relation determines the structure of a shallow bound state from the scattering length and the binding energy. Because both the scattering length and the binding energy are observable quantities, this is called model-independent determination of the structure. Recently, this approach has been applied to study the structure of hadrons~\cite{Baru:2003qq,Hyodo:2011qc,Hyodo:2013iga,Hyodo:2013nka,Hyodo:2014bda,Sekihara:2014kya,Guo:2015daa}. 

The weak-binding formula in Ref.~\cite{Weinberg:1965zz} is generalized to the case of the quasi-bound state in Ref.~\cite{Kamiya:2015aea}, by using the effective field theory. The formula relates the scattering length $a_{0}$ and the eigenenergy of the quasi-bound state $E_{QB}$ to the compositeness $X$, which is the probability of finding the two-body composite component, as
\begin{equation}
a_0
=  R \Biggl\{\frac{2X}{1+X} + {\mathcal O}\left(\left|\frac{R_{\mathrm{typ}}}{R}\right| \right) + \sqrt{\frac{\mu^{\prime 3}}{\mu^{3}}} \mathcal{O} \left( \left| \frac{l}{R} \right|^{3}\right) \Biggr\},\quad
R=\frac{1}{\sqrt{2\mu E_{QB}}} .
\end{equation}
There are two length scales. $R_{\rm typ}$ is the typical length scale of the two-body interaction, corresponding to the inverse of the momentum cutoff in the effective field theory. The scale $l=1/\sqrt{2\mu \nu}$ is determined by the energy difference $\nu$ between the threshold of the two-body channel of interest and the threshold of the decay channel. This formula is obtained by the expansion of the scattering length $a_{0}$ in powers of $1/R$. An important point here is that the coefficient of the leading order of the expansion (proportional to $R$) is solely determined by the compositeness $X$. Thus, if the magnitude of the eigenenergy of the quasibound state is small and the correction terms of ${\mathcal O}\left(\left|R_{\mathrm{typ}}/R\right| \right)$ and $\mathcal{O} \left( \left| l/R \right|^{3}\right)$ are suppressed, the compositeness $X$ can be determined only from the ratio of $a_0$ and $R$. 

The eigenenergy (pole position) of the $\Lambda(1405)$ and the scattering length of the $\bar{K}N$ channel are determined in the NLO chiral SU(3) dynamics constrained by the SIDDHARTA data. To apply this method to the $\Lambda(1405)$, we first check the size of the correction terms. Estimating the length scale $R_{\mathrm{typ}}$ and $l$ by the $\rho$ meson exchange and the energy difference of the $\pi\Sigma$ and the $\bar{K}N$ thresholds, we find $|R_{\mathrm{typ}}/R|\lesssim 0.17$ and $|l/R|^3 \lesssim 0.14$ for the eigenenergies found in chiral SU(3) dynamics, $|R| \gtrsim 1.5\hspace{1ex}\mathrm{fm}$. We thus determine the compositeness $X$ by neglecting the correction terms. The results are shown in Table~\ref{tab:Lambda}. In all cases, the compositeness $X$ and $\tilde{X}=(1-|1-X|+|X|)/2$ are found to be close to unity. Hence, the $\bar{K}N$ composite structure is the dominant component of the $\Lambda(1405)$.

\begin{table}[bt]
		\begin{tabular}{lllll}
		\hline
             Reference & $E_{QB}\physdim{MeV}$ & $a_0 \physdim{fm} $ & $X_{\bar{K}N}$ & $\tilde{X}_{\bar{K}N}$   \\  \hline
			 \cite{Ikeda:2012au}  & $-10-i26$ & $1.39 - i 0.85$ 
			 & $1.2+i0.1$ & $1.0$ \\ 
			 \cite{Mai:2012dt}  & $-\phantom{0}4-i\phantom{0}8$ & $1.81-i0.92$ 
			 & $0.6+i0.1$ & $0.6$ \\ 
			 \cite{Guo:2012vv}  & $-13-i20$ & $1.30-i0.85$ 
			 & $0.9-i0.2$ & $0.9$  \\
			 \cite{Mai:2014xna}  & $\phantom{-0}2-i10$ & $1.21-i1.47$ 
			 & $0.6+i0.0$ & $0.6$  \\ 
			 \cite{Mai:2014xna}  & $-\phantom{0}3-i12$ & $1.52-i1.85$ 
			 & $1.0+i0.5$ & $0.8$ \\ 
			 \hline
		\end{tabular} 
		\caption{Compositeness of the $\Lambda (1405)$. Shown are the eigenenergy $E_{QB}$, $\bar{K}N(I=0)$ scattering length $a_{0}$, the $\bar{K}N$ compositeness $X_{\bar{K}N}$ and $\tilde{X}_{\bar{K}N}$.}
		\label{tab:Lambda}
\end{table}

\section{SUMMARY}

The recent progress in the study of the $\Lambda(1405)$ and the $\bar{K}N$ interaction is discussed. We have established the systematic approach with the next-to-leading order chiral SU(3) dynamics together with the $K^{-}p$ scattering data and the SIDDHARTA data of the kaonic hydrogen. These enable us to determine the property of the meson-baryon scattering amplitude and the pole structure of the $\Lambda(1405)$ near the $\bar{K}N$ threshold. There are still remaining uncertainty in the $I=1$ amplitude and in the energy region far below the $\bar{K}N$ threshold. As future perspective, theoretical studies of the $\pi\Sigma$ spectra in various reactions will be important, in conjunction with the experimental developments. The studies of the wavefunction and the compositeness consistently indicate the $\bar{K}N$ molecular structure of the $\Lambda(1405)$.

\section{ACKNOWLEDGMENTS}
The author is grateful to Mike Pennington for the kind invitation to  HADRON 2015 conference and for suggesting the perfect title for the presentation. He also thank the collaborators of a series of works presented here, Yoichi Ikeda, Yuki Kamiya, Kenta Miyahara, Shota Ohnishi, Eulogio Oset, and Wolfram Weise. This work is supported in part by JSPS KAKENHI Grants No. 24740152 and by the Yukawa International Program for Quark-Hadron Sciences (YIPQS).




%

\end{document}